# Thermoelectromotive force of hafnium at plastic deformation in regime of creep at temperature of 300 K

E. V. Karaseva, P. A. Kutsenko, O. P. Ledenyov, V. I. Sokolenko, V. A. Frolov

*National Scientific Centre Kharkov Institute of Physics and Technology, Academicheskaya 1, Kharkov 61108, Ukraine.*

The thermoelectromotive force and electrical resistivity of the polycrystalline *hafnium* (*Hf*) with the grain size of *10 μm* during the process of plastic deformation in the regime of creep at the temperature of *300 K* are precisely measured. It is shown that the thermoelectromotive force depends on the deformation mechanism nature, because of the changing magnitude of the electrons scattering on the different structural defects in the *hafnium* crystal lattice at the deformation process. The main research conclusion is that the method of thermoelectromotive force measurements is more informative, comparing to the method of electric resistivity measurements in the process of accurate characterization of the *hafnium*.

## Introduction

The considerable interest to the research on the physical properties of *hafnium* (*Hf*) can be explained by the fact that this metal is one of the most perspective neutrons absorbing materials for the application in the nuclear reactors in [1]. The information on the mechanical properties of *hafnium* in the scientific literature is not complete; in particular, the important data on the physical characteristics of *hafnium* in the process of creep in a wide range of temperatures is not present. The *hafnium* experiences the long term actions by the static pressures and temperatures, hence the *hafnium* is in the process of creep during the process of operation at the nuclear reactor, that is why the accurate characterization of *hafnium* in the process of plastic deformation in the regime of creep has a significant practical meaning. Going from other point of view, the plastic deformation of *hafnium* in the regime of creep can be considered as an effective method of experimental testing, allowing to get more data on both the plastic deformation characteristics and the nature of mechanisms of plastic deformation, including the data on the structural parameters such as the dimension of grains, state of grains boundaries, density of dislocations, presence of impurities in the crystal lattice, etc.

The influences by the deformation on the electrophysical properties of the *hafnium* and other transition metals were not fully researched. The electric resistivity $\rho$, which strongly depends on the state of material structure, is frequently selected as a main electrophysical characteristic of material and usually researched in details. The thermoelectromotive force $S$ is another electrophysical characteristic of material, which exhibits a high dependence on the electronic structure of metals and alloys in [2]. In the literature, there is some data on the use of this research methodology in the experiments with the applications of external pressures in [3].

In this article, we research the possibility of application of the method of electromotive force measurements with the purpose to test the *hafnium* structural state in the process of its plastic deformation in the regime of creep. Therefore, the main purpose of our research is to find the possible correlations between the changes of magnitudes of the thermoelectromotive force, electric resistivity and velocity of *hafnium* movement in the process of plastic deformation in the regime of creep at the increasing mechanical stress pressures. Also, the clarification of possible connection of the thermoelectromotive force, electric resistivity and velocity of movement of *hafnium* in the process of creep with the characteristics of structural state of material, exposed to the plastic deformation.

## Hafnium Samples Synthesis

We have researched the samples of *hafnium* of the type of *ГФЭ-1* with the purity of *99,7 mass %*, which were annealed at the temperature of *1173 K* over the time period of *1* hour, and had the impurities contents *O < 0,05 mass %*, *Fe ≈ 0,03…0,04 mass %*, *Zr ≈ 0,2 mass %*, and the average grain dimension of *10 μm* in the *hafnium* crystal lattice The researched samples of *hafnium* had the geometric shapes of parallelepiped with the dimensions of *0,2×4×65 mm*.

The differential thermoelectromotive force is

$$S = \frac{\Delta U(\Delta T)}{\Delta T},$$

where *ΔT* is the difference of temperatures between the electric potentials contacts, which was determined in relation to the *copper*; *ΔU(ΔT)* is the difference of electric potentials, appearing as a result of the presence of gradient of temperatures of *ΔT/l*, were $l \approx 30\ mm$ is the distance between the electrodes, measured by the potentiometer of the model of *P363-3*. The electric potentials contacts, which were symmetrically placed in relation to the edges of a sample, were situated in the zone of deformation outside the mechanical captures. The place of electric contact between the electric potentials contacts and the sample was made of the thin plastic interlayer of *indium*, which was attached to the sample by the adhesion method with the purpose to preserve the initial structure of a sample in the place of electric contact without the possible distortion of measured results. The electric heater was attached to one of the edges of a sample with the aim to create a gradient of temperatures. The difference of temperatures *ΔT* was measured, using the differential method with the help of the two serially connected thermo-pairs made of the *copper – constantan* with the electrically isolated solders, which were fixed on the plane of a sample, which is opposite to the plane with the electric potentials contacts. The accuracy of temperature measurements was ± *0,1 K*. The magnitudes of the thermoelectromotive force *S*, velocity of movement *ε*, electric resistivity *ρ* at the plastic deformation of *hafnium* in the regime of creep were simultaneously measured by the validated experimental methods. The deformation of samples of *hafnium* was conducted, adding the mechanical stress pressures in the steps with the increased magnitude of deformation stress $\sigma \approx 5\ MPa$ per every step. The accuracy of measured values of added length *Δl* of a sample was $5\cdot10^{-5}\ cm$. The total error magnitude was not more than *0,05 %* at the electric resistivity *ρ* measurements by the potentiometer of the model of *P363-3*, applying the four points scheme. The deviation of measured magnitudes of electric resistivity *ρ* was equal to ± *0,5 %*.

## Experimental Measurements Results

In Fig. 1, the obtained experimental measurements results on accurate characterization of physical parameters of *hafnium* are shown, including:
**a)** Dependence of the velocity of movement of *hafnium* on the real mechanical stress pressure of *hafnium* *ε̇ (σ)* in the process of plastic deformation of *hafnium* in the regime of creep at the temperature of *300 K*;
**b)** Dependence of the normalized electric resistivity of *hafnium* on the real mechanical stress pressure of *hafnium* *ρ/ρ₀ (σ)* in the process of plastic deformation of *hafnium* in the regime of creep at the temperature of *300 K* ($\rho_0$ is the electric resistivity of *hafnium* without the mechanical stress pressure);
**c)** Dependence of the normalized thermoelectromotive force of *hafnium* on the real mechanical stress pressure of *hafnium* $S/S_0\ (\sigma)$ in the process of plastic deformation of *hafnium* in the regime of creep at the temperature of *300 K* ($S_0$ is the thermoelectromotive force of *hafnium* of *hafnium* without the mechanical stress pressure).

As it can be seen in the graphics, all the three measured characteristics of *hafnium* non-monotonically change at the increase of the mechanical stress power in the process of plastic deformation in the regime of creep.

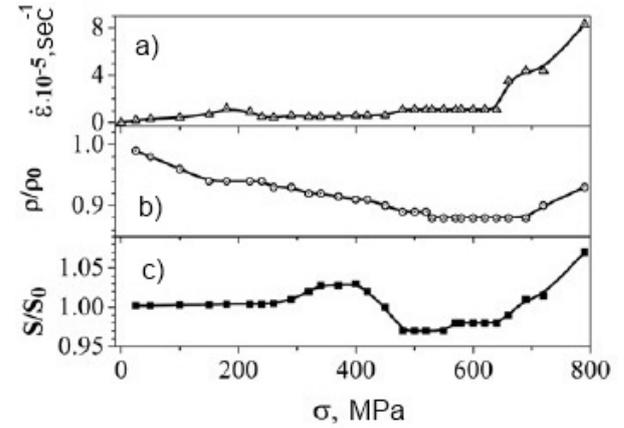

*Fig. 1. Experimental measurements results:*
*a) Dependence of velocity of movement of hafnium on real mechanical stress pressure of hafnium ε̇ (σ) in process of plastic deformation of hafnium in regime of creep at temperature of 300 K;*
*b) Dependence of normalized electric resistivity of hafnium on real mechanical stress pressure of hafnium $\rho/\rho_0(\sigma)$ in process of plastic deformation of hafnium in regime of creep at temperature of 300 K ($\rho_0$ is the electric resistivity of hafnium without the mechanical stress pressure);*
*c) Dependence of normalized thermoelectromotive force of hafnium on real mechanical stress pressure of hafnium $S/S_0\ (\sigma)$ in process of plastic deformation of hafnium in regime of creep at temperature of 300 K ($S_0$ is the thermoelectromotive force of hafnium of hafnium without the mechanical stress pressure).*

There are the two intervals of mechanical stress pressure: 1) *σ < 650 MPa* and 2) *650 < σ ≤ 800 MPa*, at which the velocities of movement of *hafnium* significantly differ in the process of plastic deformation in the regime of creep. The velocity of movement increases up to $\varepsilon \approx 1,5\cdot10^{-5}\ sec^{-1}$ at the increase of mechanical stress pressure up to $\sigma \approx 650\ MPa$; then the velocity of movement *ε* sharply increases at *650 < σ ≤ 800 MPa*.

The non-monotonic change of the velocity of movement of *hafnium* at the magnitude of mechanical stress pressure below the limit of fluidity *σ < 450 MPa* can be connected with the process of re-distribution or exhaustion of dislocations in the crystal lattice of *hafnium* [4]. In particular, the small maximum of the velocity of movement of *hafnium* at the mechanical stress pressure $\sigma \approx 200\ MPa$ can be connected with the disappearance of weakly anchored dislocations, because of not enough annealing of a sample. At this process, the small increase of the magnitude of thermoelectromotive force and the big decrease of electric resistivity (~*10 %*) are observed. The constant decrease of electric resistivity in this range of mechanical stress pressure confirms the supposition that there are the decrease of a number of defects and the decrease of level of internal stresses in the crystal lattice structure of *hafnium*. The interval with the small increase of the velocity of movement $\dot{\varepsilon}$ corresponds to the interval with the decrease of the normalized electric resistivity $\rho/\rho_0$.

The significantly non-monotonous character of dependence of the normalized electromotive force on the mechanical stress pressure $S/S_0$ ($\sigma$) at $\sigma \approx 300...500$ *MPa* attracts a considerable attention. The strong non-monotonous changes of normalized thermoelectromotive force have place in the region of mechanical stress pressures below the limit of fluidity (*300...400 MPa*), where the velocity of movements at plastic deformation in the regime of creep is practically not changing. Also, it is possible to note that in the region of the mechanical stress pressures $\sigma \approx 500...650$ *MPa*, the step on the curve in the dependence $S/S_0=f(\sigma)$ is observed, but the velocity of movement and the electric resistivity remain constant.

At the second interval of *650 < σ ≤ 800 MPa*, all the researched dependences are well correlated. In the frames of the existing representations, the decrease of electric resistivity in the process of plastic deformation in the regime of creep is also connected with the re-distribution of dislocations [5]. As it was shown in [4], the plastic movement of *hafnium* at the temperature of *300 K* and at the mechanical stress pressures above the limit of fluidity depends on the influence by the following processes: 1) the inter-grain sliding, controlled by the dislocations, and 2) the inter-grain sliding along the grain boundaries. It is possible to assume that, in the researched case in the conditions of creep, there are similar physical processes, hence the significantly non-monotonous character in the dependence $S/S_0$ ($\sigma$) provides the data about the physical properties of an interconnection between the thermoelectromotive force and the structural changes in the crystal lattice of *hafnium*. The contribution by every physical mechanism is not constant, because it depends on the applied mechanical stress pressure. The steps on the curves in the dependence of the velocity of movement on the mechanical stress pressure $S/S_0$ ($\sigma$) correspond to the magnitudes of mechanical stress pressure at which the dominant deformation mechanisms change. The similar steps are observed on the dependences $\dot{\varepsilon} = f(\sigma)$ and $S/S_0=f(\sigma)$. In the conditions of experiment, the thermoelectromotive force and the electric resistivity were determined by the processes of electron scattering on the phonons ($\rho_T$), impurities ($\rho_1$), structural defects ($\rho_2$), etc. The contribution by every described scattering mechanism changes during the process of plastic deformation [6].

In agreement with the *Mott, Jones* theory [7, 8], the thermoelectromotive force and the electric resistivity are interconnected parameters, hence the general expression for the electromotive force $S$ can be presented as in eq. (1)

$$S = -f(T)\left[\frac{\rho_T}{\rho}\frac{d\ln\rho_T}{d\ln E} + \frac{\rho_1}{\rho}\frac{d\ln\rho_1}{d\ln E} + \frac{\rho_2}{\rho}\frac{d\ln\rho_2}{d\ln E} + ...\right], \quad (1)$$

where $f(T)$ is some general function of the certain parameters on the temperature; $E$ is the characteristic energy of electrons.

In the frames of the two zones model of the transition metal, the magnitude of the thermoelectromotive force can be expressed as in eq. (2) in agreement with the *Mott, Jones* theory [7]

$$S = \frac{\pi^2 k_B^2 T}{3e}\left[\frac{1}{N_d}\frac{\partial N_d}{\partial E} - \frac{1}{F}\frac{\partial F}{\partial E}\right]_{E=E_F}, \quad (2)$$

where $k_B$ is the *Boltzmann* constant; $T$ is the temperature; $e$ is the electron charge; $F$ is the area of *Fermi* surface; $N_d$ is the density of states in the *d*-zone.

In [9, 10], the analysis of temperature dependences of the electric resistivities of transition metals with the deformation and tempering defects and impurities was completed in the frames of the two zones model. In [9, 10], it is shown that the increase of the density of dislocations results in the narrowing of *d*-zone, increase of the density of states of *d*-electrons and enhancement of anisotropy of the *Fermi* surface (the effective decrease of curvature of the *Fermi* surface). The observed change of sign of gain of the ratio $S/S_0$ at *300 < σ ≤ 500 MPa* can be connected with the change of the relation between the first and second composed parts in the formula (2), because of the differing changes of density of states and characteristics of the *Fermi* surface at the various stages of change of the *hafnium* structure in the process of plastic deformation in the regime of creep.

The local electron energy $E$ in the material depends on both the elastic deformations of crystal lattice and the degree of material deformation. The scattering of electrons, determining the magnitude of electric resistivity, depends on the competition between the physical processes, including the transformation, re-distribution, and disappearance of the centers of scattering of various types, which originate at the change of structural state of *hafnium* in the result of plastic deformation of a sample. Therefore, it is necessary to have the data about the change of structure with the re-distribution of defects at various stages of the plastic deformation process in the regime of creep to be able to interpret the obtained dependence $S/S_0 = f(\sigma)$.

Let us estimate the magnitude of energy change, which occurs in the electron sub-system in the process of plastic deformation, considering the curve in Fig. 1 (b). In the case of metals, the quantum theory gives the following expression in [11]

$$S = \frac{\pi^2 k_B^2 T}{3e} \left[ \frac{1}{\mu} + \frac{1}{l} \frac{dl}{dE} \right],$$

where $\mu$ is the chemical potential of electrons, which is equal to the *Fermi* energy in the case of metals; $l$ is the mean free path of electrons with the kinetic energy $E$. In the metals, it is possible to assume that $l \sim E^2$, and taking to the consideration the fact that the electrons with $E \approx \mu$ take part in the thermal drift, then we can write the following expression

$$\frac{1}{\mu} + \frac{1}{l} \frac{dl}{dE} = \frac{3}{\mu},$$

hence

$$S = \frac{\pi^2 k_B^2 T}{e\mu}, \text{ and } S = \frac{T}{\mu}.$$

Thus, the change of the *Fermi* level of electrons $\mu$ has a main influence on the magnitude of the thermoelectromotive force in the process of plastic deformation of *hafnium* in the regime of creep at the given temperature and the constant gradient of temperatures between the edges of a sample. It is possible to assume that the structural defects, appearing in the crystal lattice of a sample of *hafnium* at the different stages of the plastic deformation process, play an important role in this process.

As it can be seen in Fig. 1(**c**), the deviations of the magnitude of electromotive force $S$ at the mechanical stress pressure from the magnitude of electromotive force $S_0$ without the mechanical stress pressure is approximately equal to ± *(3…4) %*, therefore, it is possible to assume that the deviations of the chemical potential of electrons $\mu$ at the mechanical stress pressure from the chemical potential of electrons $\mu_0$ without the mechanical stress pressure must have the same value, that is quite realistic assumption, taking to the consideration the estimation of possible influence by the expanding pressure on the energy state of electron system. It is a well known fact that the characteristic electron pressure in the metals in by the *Fermi* statistics of electrons has the magnitude $P_F \approx 10^{10} - 10^4$ *MPa*. Therefore, the relation of the mechanical stress load to the characteristic electron pressure is equal to $\sigma/P_F \approx$ *(3…5) %* at *300 < σ ≤ 500 MPa*. The proximity of this estimation to the observed changes of $S/S_0$ especially at *σ > 500 MPa* can be considered as an evidence that there is a connection of the given thermoelectric coefficient with the denoted theoretical mechanism.

The obtained experimental results confirm our assumption that the physical behavior of the thermoelectromotive force is influenced by both *1)* the various plastic deformation mechanisms, which are related to the changing velocity of movement of *hafnium* in the process of plastic deformation in the regime of creep, and *2)* the unique physical processes, which have place at the electron sub-system. We believe that this is a most interesting research result, which needs to be further investigated.

# Conclusion

In this research, the thermoelectromotive force and electric resistivity of the polycrystalline *hafnium* with the grain size of *10 μm* at the process of plastic deformation in the regime of creep at the temperature of *300 K* are precisely measured. It is shown that the thermoelectromotive force depends on the deformation mechanism nature, because of the changing magnitude of the electrons scattering on the different structural defects in the *hafnium* crystal lattice at the plastic deformation process. It is shown that, in the process of one axis stretching, the change of the thermoelectromotive force of the polycrystalline *hafnium* with the mean grain dimension of *10 μm* is well correlated with the changes of the electric resistivity and velocity of movement of *hafnium* in the process of plastic deformation in the regime of creep at the increasing stress load *(650 < σ ≤ 800 MPa)*. It is possible to make a conclusion that the measurement method of the magnitude of the thermoelectromotive force $S$ is well suited for the accurate characterization of the structural state of a sample in the given magnitudes of mechanical stress pressure. There is a well identified maximum on the dependence of the thermoelectromotive force on the mechanical stress pressure $S/S_0 = f(\sigma)$, which can be observed up to the limit of plastic deformation at *300 < σ ≤ 500 MPa*; there are no similar maximums on the dependences: $\rho/\rho_0 = f(\sigma)$ and $\acute{\varepsilon} = f(\sigma)$. This fact confirms our assumption that it is possible to make the accurate characterization of the structural and electron states of a sample of *hafnium* under the mechanical stress pressure, using the discussed research method. We think that the observed physical behavior of thermoelectromotive force of *hafnium* in the process of creep must be further researched experimentally and theoretically. In our opinion, the method of thermoelectromotive force measurements is more informative, comparing to the method of electrical resistivity measurements in the process of accurate characterization of the *hafnium* crystal lattice parameters, because the resolution of the method of thermoelectromotive force measurements is much higher and there is no need to consider the ongoing changes of the length and cross-section of a sample.

This experimental research is completed in the frames of the nuclear science and technology fundamental research program at *the Schubnikov Cryogenic Laboratory at the National Scientific Centre Kharkov Institute of Physics and Technology* (*NSC KIPT*) in Kharkov in Ukraine. The research was funded by *the National Academy of Sciences* in Ukraine.

This research paper was published in the *Problems of Atomic Science and Technology* (*VANT*) in 2011 in [12].

*E-mail:   ledenyov@kipt.kharkov.ua